\documentclass[aip,jcp,amsmath,amssymb,floatfix,reprint,twocolumn]{revtex4-1}
\usepackage{amssymb}
\usepackage{amsmath}
\usepackage{graphicx}
\usepackage{amsbsy}
\usepackage{color}
\usepackage{bm}
\usepackage{epsfig}
\usepackage{graphicx}
\usepackage{epstopdf}
\usepackage{tikz}
\usepackage{setspace}
\usepackage{tabularx}
\usepackage{booktabs}
\usepackage{siunitx}
\usepackage{braket}
\usepackage{pgfplots}

\newcommand{\bR}{\mathbf{R}}

\newcommand{\bOm}{\mathbf{\Omega}}

\begin{document}
\title{Left or right cholesterics? A matter of helix handedness and curliness}

\author{Elisa Frezza} 
\altaffiliation{Current address: Bases Mol\'{e}culaires et Structurales des Syst\`{e}mes Infectieux, Univ. Lyon I / CNRS UMR 5086, IBCP, 7 passage du Vercors, 69367 Lyon, France.} 
\affiliation{Dipartimento di Scienze Chimiche, 
Universit\`{a} di Padova, via F. Marzolo 1, 35131 Padova, Italy}

\author{Alberta Ferrarini} 
\email{alberta.ferrarini@unipd.it}
\affiliation{Dipartimento di Scienze Chimiche, 
Universit\`{a} di Padova, via F. Marzolo 1, 35131 Padova, Italy}

\author{Hima Bindu Kolli} 
\affiliation{Dipartimento di Scienze Molecolari e Nanosistemi, 
Universit\`{a} Ca' Foscari di Venezia,
Dorsoduro 2137, 30123 Venezia, Italy}

\author{Achille Giacometti} 
\email{achille.giacometti@unive.it}
\affiliation{Dipartimento di Scienze Molecolari e Nanosistemi, 
Universit\`{a} Ca' Foscari di Venezia,
Dorsoduro 2137, 30123 Venezia, Italy}

\author{Giorgio Cinacchi} 
\email{giorgio.cinacchi@uam.es}
\affiliation{
Departamento de F\'{i}sica Te\'{o}rica de la Materia Condensada and
Instituto de F\'{i}sica de la Materia Condensada, 
Universidad Aut\'{o}noma de Madrid,
Campus de Cantoblanco, 28049 Madrid, Spain
}

\date{\today}

\begin{abstract}
\noindent Using an Onsager-like theory, we have investigated the relationship between the morphology of hard helical particles  and the features (pitch and handedness) of the cholesteric phase that they form.  
We show that right-handed helices can assemble into right- ($\cal R$) and left-handed ($\cal L$) cholesterics, depending on their curliness, and that the cholesteric pitch is a non-monotonic function of the intrinsic pitch of particles.
The theory leads to the definition of a hierarchy of pseudoscalars, 
which quantify the difference in the average excluded volume between
pair configurations of helices having ($\cal R$) and  ($\cal L$)-skewed axes.
The predictions of the Onsager-like theory are supported by Monte Carlo simulations of the isotropic phase of hard helices,
showing how the cholesteric organization, which develops on scales longer than hundreds of molecular sizes, is encoded in the short-range chiral correlations between the helical axes.
\end{abstract}
\maketitle
\section{Introduction}
\noindent Recently, we have undertaken a thorough investigation of the phase diagram of hard helices, aimed at elucidating its general features and in particular its dependence upon the helix morphology.\cite{JCPI,JCPII} To this purpose we  are combining   
classical density functional theory and numerical simulations. We have found that the phase diagram is strongly different from that of hard spherocylinders, to which helices are often assimilated, and shows new phases that are special to helices. The present work focusses on the cholesteric phase (N$^\ast$), which is a chiral nematic phase where the director rotates in helical way around a perpendicular axis. It is characterized by the pitch $\cal P$, or the corresponding wavenumber $q=2\pi/{\cal P}$.  Conventionally, these are taken as positive for a right-handed ($\cal R$) helix, and negative for a left-handed ($\cal L$) helix.  
Typical values of the pitch range from hundreds of nanometers for strongly chiral molecular species to hundreds of micrometers for chiral colloidal systems, which are well beyond the length scale of inter-particle interactions.\cite{note}

The specific question that we want to address here is the relationship between the morphology of the helical particles and the features (handedness and pitch) of their cholesteric phase.

\begin{figure}[h]
\begin{center}
\includegraphics[scale=0.25]{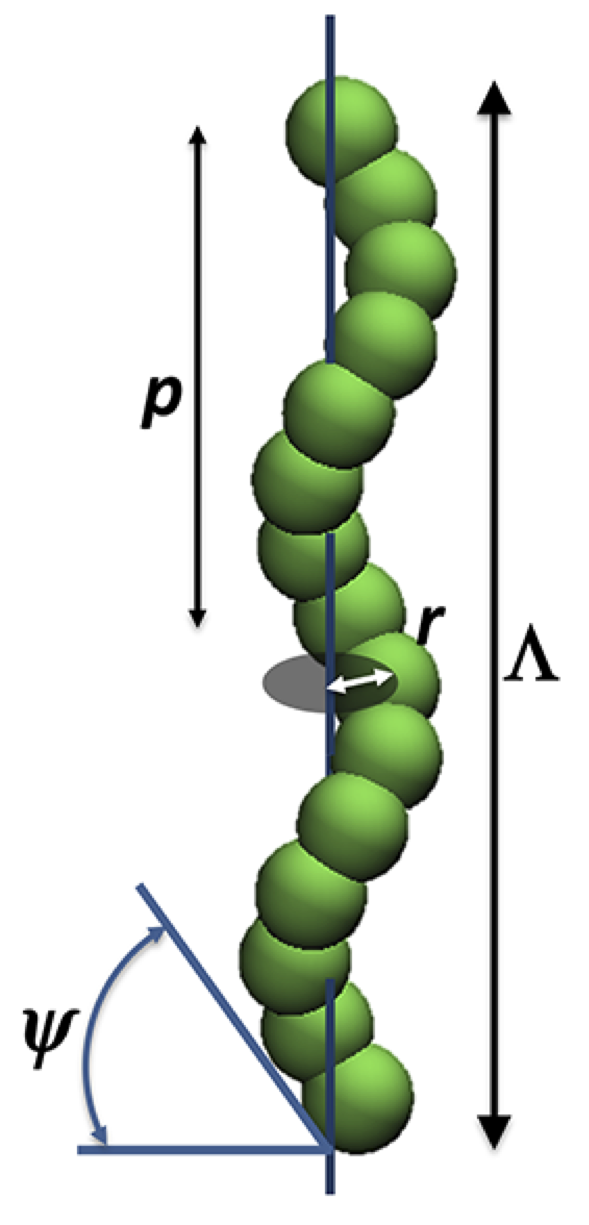}
\caption {A model helix and its characteristic parameters: $r$ is the radius, $p$ is the pitch, $L$ is the contour length, $\Lambda$ is the Euclidean length and $\psi$ is the inclination angle.}
\label{fig:helix}
\end{center}
\end{figure}

That hard threaded helices can form a cholesteric phase to optimize their packing was recognized by the early Straley work.\cite{Straley} 
Best packing of right-handed screws is obtained for  either $\cal R$ or $\cal L$-skewed configurations,
in the case of tightly and loosely twisted threads, respectively.
Thus, based on simple geometry considerations, the cholesteric handedness is expected to depend not only on the handedness of the constituent helices, but also on their specific geometry features.\cite{Harris1999,Cherstvy2008} 
Taking as an indicator the inclination angle $\psi$ between the tangent to helix at a given point and the plane normal to the helix axis passing through the same point (see  Fig. \ref{fig:helix}), the phase chirality is expected to 
be the same as that of the constituent helices for $\psi < 45^\circ$ 
and the opposite for $\psi > 45^\circ$. 

Another issue concerns the relationship between the geometry of the helical particles and the magnitude of the cholesteric  pitch. 
Theoretical predictions are somehow controversial, which may also depend on the assumptions adopted to derive expressions for the cholesteric pitch. For rigid threaded rods,  $\cal P$ independent of the intrinsic pitch $p$ of particles was predicted by Straley. \cite{Straley} Subsequently, a more general relationship was obtained for the same kind of particles, which for small intrinsic pitch   reduces to ${\cal P} \propto p^{-1}$, in agreement with intuitive arguments.\cite{Pelcovits1996}
On the other hand, ${\cal P} \propto p$ was predicted for hard twisted rigid ellipsoids.\cite{Evans1992}

Clarifying the relationship between the handedness and pitch of the cholesteric phase and the chirality of the constituent particles has been a long-standing goal.\cite{Harris1999} Over the years our understanding has grown,
\cite{Harris1997,JacksonJMolPhys2011,KatsonisJMaterChem} and for some kinds of systems, like for instance low molecular weight thermotropic liquid crystals,  predictive models are available.\cite{Pieraccini11}  A general result is that the cholesteric organization is determined by a subtle balance of intermolecular interactions, which promote director twists in opposite sense. 
Therefore a detailed knowledge of the system is a necessary requirement for predictions that can be compared with experimental data. The behaviour of helical polymers is difficult to rationalize, for two main reasons related to the intricacy of polymer-polymer 
interactions.\cite{Sato1998,LivolantDNA,Zugenmaier} These are strongly dependent on the polymer conformation, which includes not only the backbone structure, but also the conformation of sidechains.   Moreover, in lyotropic systems the inter-polymer interactions are mediated by  solvent and any species that is present, including ions and small molecules.
Molecular theories for the cholesteric phase of polymers
have been presented in the literature, where
electrostatic or dispersion interactions with helical symmetry were superimposed to steric repulsions between either hard rigid rods \cite{wensink_JPhysCondensMatter,wensink:234911} or hard rigid helices.\cite{Osipov85,Sato1998,Emelyanenko2003,Tombolato05,Frezza:SoftMatter} 
Here we will focus on rigid helices with pure hard-core interactions, where the twist of the nematic director is due to packing reasons only.
This allows us to highlight some general aspects of the relationship between helical shape and cholesteric organization, and can represent a useful reference for understanding the behavior of more complex systems.
In particular, we will examine  the same model of rigid hard helices used in our previous studies (see  Fig. \ref{fig:helix})\cite{JCPI,JCPII} which, unlike the threaded rods generally considered so far,\cite{Straley} can be highly non-convex in shape. 

The next section gives an outline of the Onsager-like theory for the cholesteric phase and introduces a set of pseudoscalar parameters that quantify the chirality of the excluded volume between pairs of helical particles.  
Then, we will report and discuss the  results of calculations for helices having different structural parameters. In the same section we will present a comparison with suitable correlation functions, obtained from Monte Carlo simulations of the isotropic phase of the same kind of helical particles. Finally we will draw the conclusions of our study.

\section{Theoretical background}
\noindent The same Onsager-like approach as in ref. \cite{JCPI} has been used. Here, a form suitable for the cholesteric phase is adopted, as in refs. \cite{Tombolato05,Frezza:SoftMatter}. Only the main points of the theory will be presented; details can be found in the previous references. 
The Helmholtz free energy of a system of $N$ identical hard helices in the volume $V$ (or volume per particle $V/N=v$) at the temperature $T$ is expressed as a functional of the single particle density function $\rho(\bR,\bOm)$, where $\bR$ is the particle position and $\mathbf{\Omega}=(\alpha,\beta,\gamma)$ are  the Euler angles specifying the particle orientation in a laboratory (LAB) frame, with  the normalization condition $\int d \bR d\bOm \rho(\bR,\bOm)=N$. 
In a laboratory (LAB) frame with the $Y$-axis parallel to the cholesteric axis the density function can be expressed  $\rho(\bR,\bOm)=\rho f(\bOm|Y)$ with $\rho=N/V$ and the orientational distribution function $f(\bOm|Y)$, whose position dependence simply reflects the helical periodicity of the director. This is normalized as $\int d\bOm f(\bOm)=1$, irrespective of $Y$.

The ideal contribution to the Helmholtz free energy is given by:
\begin{equation}
A^\text{id}= Nk_BT\left[\ln \left (\frac{\Lambda_{\text{tr}}^3}{V} \frac{\Theta_{\text{or}}}{T} \right ) + \ln N -1 \right] +A^\text{or}
\label{eq:A}
\end{equation}
where the first term is the Helmholtz free energy of the ideal gas:\cite{McQuarrie}
$\Lambda_{\text{tr}}=\left(h^2 / 2 \pi k_B T m \right)^{1/2}$  is the de Broglie wavelength and 
$\Theta_{\text{or}}=h^2/8\pi^2k_B I$ is the rotational temperature, with $k_B$ and $h$ being the Boltzmann and the Planck constant, respectively, while $m$ is the mass and $I$ is the inertia moment of a particle.
The last term on the rhs of Eq. \eqref{eq:A} accounts for the decrease in entropy due to orientational ordering: 
\begin{equation}
\frac{A^\text{or}}{Nk_BT}=
 \int  d\bOm f(\bOm) \ln \left [8\pi^2  f(\bOm) \right ]
\label{eq:AorSM}
\end{equation}
where again no $Y$-dependence appears, since the integral over all orientations is independent of the position. 
Within the Onsager-like theory a second virial approximation of  
the excess free energy  is used,
which is expressed in terms of the interaction between a pair of particles (1, 2).
Considering a LAB frame with its origin at the position of one particle (1), we can write: 
\begin{equation}
\begin{split}
\frac{A^\text{ex}}{Nk_B T}=&-G(\phi) \frac{\rho}{2}\int d\bOm_{1} f(\bOm_1|0)  \int d\bOm_{2} \\
& \int d\bR_{12} f(\bOm_2|Y_{12}) e_{12}(\bR_{12},\bOm_{12})
\end{split}
\label{eq:aexSM}
\end{equation}
where 
$\bOm_{12}$ are the Euler angles specifying the rotation from particle 1 to particle 2, and  
$\bR_{12}=\bR_{2}-\bR_{1}$.
Here $e_{12}$ is the Mayer function,\cite{McQuarrie} which in the case of purely hard-core interactions is equal to -1 for overlapping pair configurations and vanishes for any other configuration.  
The  pre-factor $G(\phi)$ is the so-called Parsons-Lee correction 
$G(\phi)=(1/4)(4-3\phi)/(1-\phi)^2$
where $\phi=v_0/v$ is the packing fraction.
For non-convex particles a variant has been proposed (modified Parsons-Lee) wherein the geometric volume $v_0$ is replaced by  an effective volume, $v_{\text{ef}}$, defined as the volume of a particle that is inaccessible to other particles.\cite{Varga2000} 
 
The excess free energy has an intrinsic dependence upon the twist deformation of the director, which can be parameterized by the wave vector $q$. For deformations on a scale much  longer than that of intermolecular interactions, as in the cholesteric phase, 
this dependence can be approximated by the truncated Taylor series expansion: 
\begin{equation}
a^\text{ex}=a^\text{ex}|_{q=0}+q \frac{\partial a^\text{ex}}{\partial q}|_{q=0}+\frac{1}{2} q^2 \frac{\partial^2 a^\text{ex}}{\partial q^2}|_{q=0}
\label{eq:aexp}
\end{equation}
The first term is the free energy density of the undeformed nematic phase:
\begin{equation}
a^\text{ex}_u \!  = \!  k_B T G(\phi) \dfrac{\rho^2}{2}  \!  \int \!  \!  \!  d\bOm_{1} f(\bOm_1) \! \!  \!  \int \!  d\bOm_{2}   f(\bOm_2) v_{excl} (\bOm_{12})
\label{eq:aex0}
\end{equation}
with $v_{excl} (\bOm_{12})=- \int d \bR_{12}\, e_{12}(\bR_{12},\bOm_{12}) $
being the volume excluded to particle 2 by particle 1, when they are in the relative orientation defined by the angles $\bOm_{12}$.
Here $f(\bOm_i)$  is the orientational distribution function in the undeformed nematic phase. 
The local order is assumed to be unaffected by long wavelength twist distortions;
thus, $f(\bOm_i)$ in the N* phase has the same form as in the undeformed N phase, but with respect to a director that rotates with the $Y$ coordinate. 
If the orientational distribution function of particle 2 is  expressed as $f(\bOm_2,\chi)$, where $\chi=qY_{12}$ is the angle between the director at the position of particle 2 and the director at the origin of the reference frame,
the following expressions are obtained for the coefficients of the first ($i=1$) and second ($i=2$) power of $q$ in Eq. \ref{eq:aexp}:
\begin{equation}
\begin{split}
\frac{\partial^i a^\text{ex}}{\partial q^i}|_{q=0}
=  & k_B T G(\phi)  \frac{\rho^2}{2} \int d\bOm_{1} f(\bOm_1)  \\
& \int \!  \!  d\bOm_{2}  \left( \frac{\partial^i f}{\partial \chi^i} \right)_{\chi=0}  \!  \!  \!  \!  \!  f(\bOm_2) \!  \int_{v_{excl}  (\bOm_{12})} \!  \!  \!   \!  \!  \!  d\bR_{12}  Y_{12}^i 
\end{split}
\label{eq:Ai}
\end{equation}
where the inner integrals are over the excluded volume.
Eq. \ref{eq:aexp}  has the usual form of continuum elastic theory,\cite{Vertogen} where 
the coefficients of the expansion are denoted as \emph{chiral strength}, $k_2=({\partial a}/{\partial q})_{q=0}$,  and \emph{twist elastic constant}, $K_{22}= ({\partial^2 a}/{\partial q^2})_{q=0}$, respectively. 
It may be worth remarking that the terms in Eq. \ref{eq:aexp} behave differently under inversion: $a|_{q=0}$ and $K_{22}$ are invariant (they are scalar), and the chiral strength $k_2$ changes its sign (it is a pseudoscalar). Thus,  for chiral particles that are the mirror image of each other (enantiomers), the free energy of the undeformed nematic phase and the twist elastic constant are identical, whereas $k_2$ takes opposite values.
Within the Onsager-like approach, the chiral strength  is proportional to the change in the average excluded volume caused by a twist deformation of the director. A positive value, $k_2 >0$, means that the excluded volume increases for an $\cal R$ twist ($q>0$); in this case,  an  $\cal L$ cholesteric phase is favored, since this leads to a decrease of the free energy. 
Conversely, if $k_2 <0$ an  $\cal R$ cholesteric phase is formed.
The equilibrium wavenumber of the twist deformation, obtained by minimization of the free energy,  is given by: $q=-k_2/K_{22}$. It is not affected by the $G(\phi)$ factor and is indirectly affected by the density, through the orientational distribution function. 

\begin{figure}[h]
\begin{center}
\includegraphics[scale=0.3]{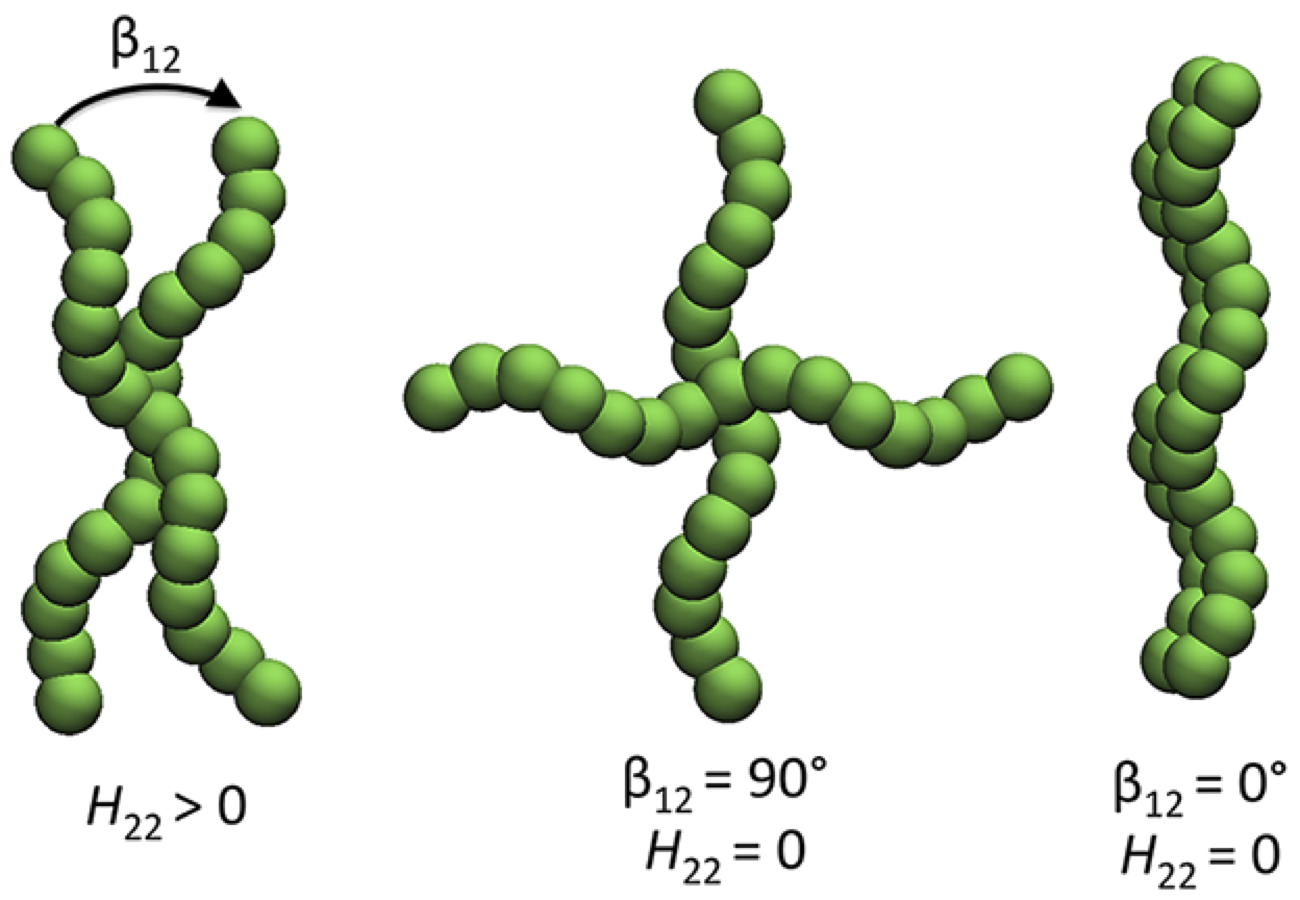}
\caption {Pair configurations of helices with different values of the inter-helix angle $\beta_{12}$ (in all cases 
$\mathbf{\hat{w}}_{12} // \hat{\textbf{R}}_{12}$). }
\label{fig:descriptor}
\end{center}
\end{figure}

\subsection{Chiral strength and chiral descriptors}
\noindent For calculations it is expedient to expand the density function on a basis of Wigner matrices,\cite{Varshalovich} which leads to expression of the free energy, chiral strength and twist elastic constant in terms of  orientational order parameters. 
After some algebraic manipulation the chiral strength, Eq. \ref{eq:Ai}  with $i=1$, 
can be expressed as the following  summation:\cite{Tombolato05} 
 \begin{equation}
 \begin{split}
k_{2}   = &  \frac{1}{3 \sqrt{2}}    \frac{k_BT}{8 \pi^2 v^2} G(\phi) \\   
& \sum_{j=0,2,4,\ldots} (2j+1) \braket{P_{j}}   \sum_{j^\prime=0,2,4,\ldots}  (2j^\prime+1)   \\   
& \sqrt{\frac{(j^\prime+1)!}{(j^\prime-1)!}}     C^2(j,j^\prime,1;0,1,1)  \braket{P_{j^\prime}} h_{j^\prime j^\prime}
\end{split}
\label{eq:k2}
\end{equation}
where $\braket{P_{j}}$ are the usual orientational order parameters, defined as the orientational average of Legendre polynomials of $j$-rank;  $C(\ldots)$ are Clebsch-Gordan coefficients and $h_{jj}$ denote the integrals:
 \begin{equation}
 \begin{split}
 h_{jj}  = & \int d\Omega_{12}\, \int_{v_{excl} (\bOm_{12})}  \!  \!  \!    \!  \!  \!  \!   \!  \!  \!  \!  \! \! d\textbf{R}_{12}\, 
d_{10}^{j}(\beta_{12}) \\
& (X_{12} \sin \alpha_{12}-Y_{12} \cos \alpha_{12}) \quad \qquad j=2,4,... 
\end{split}
\label{eq:hii}
\end{equation}
with $d_{10}^j$ being components of reduced Wigner matrices. 
The functions under the integrals, 
$H_{jj}=d_{10}^{j}(\beta_{12})  (X_{12} \sin \alpha_{12}-Y_{12} \cos \alpha_{12})/\textbf{R}_{12} $,  
can be rewritten as:
\begin{subequations} \label{eq:subeqns}
 \begin{align}
 H_{22}=&\sqrt{\frac{3}{2}} (\mathbf{\hat{u}}_1 \cdot \mathbf{\hat{u}}_2)
 (\mathbf{\hat{u}}_1 \times \mathbf{\hat{u}}_2) \cdot \hat{\textbf{R}}_{12}   \label{eq:subeqn1}\\
 H_{44}=&\frac {\sqrt{5}} {4}
 \left [ 7 (\mathbf{\hat{u}}_1 \cdot \mathbf{\hat{u}}_2)^3-3 (\mathbf{\hat{u}}_1 \cdot \mathbf{\hat{u}}_2) \right ]
 (\mathbf{\hat{u}}_1 \times \mathbf{\hat{u}}_2) \cdot \hat{\textbf{R}}_{12} \label{eq:subeqn2} \\
 \ldots  & \nonumber 
 \end{align}
 \end{subequations}
where $\mathbf{\hat{u}}_1$ and $\mathbf{\hat{u}}_2$ are unit vectors parallel to the  axes of the helical particles and
$\hat{\textbf{R}}_{12}=\textbf{R}_{12}/{R}_{12}$. 
It can be recognized that the pseudoscalars Eqs. \ref{eq:subeqns} correspond to a subset of the rotational invariant functions introduced by Stone.\cite{Stone78}

Th $h_{jj}$ integrals represent the decomposition of the excluded volume in chiral contributions of different ranks.
They can be related to the difference in excluded volume between configurations of pairs of helices having $\cal R$- and $\cal L$-skewed long axes. 
For achiral particles, e.g. for cylinders,  oppositely skewed configurations give equivalent contributions of opposite sign, so the integrals vanish. However for chiral particles, and in our specific case for helices,  this symmetry is broken and $h_{jj}\neq 0$. 
If the two chiral molecules are replaced by their mirror images (enantiomers),  
$h_{jj}$ reverses its sign.
Thus, the $h_{jj}$ pseudoscalars can be taken as simple geometric descriptors, to correlate the cholesteric pitch and handedness to the structure of the constituent particles, through their excluded volume.\cite{RossiECLC2013,FrezzaPhD}
The  lowest rank term, $h_{22}$, which generally prevails over the others, has a particularly simple geometrical interpretation.
Its integrand can be rewritten as  $H_{22}=\sqrt{3/8} (\mathbf{\hat{w}}_{12} \cdot \hat{\textbf{R}}_{12})  \sin (2 \beta_{12}) $,
where $\beta_{12}$ is the angle between them and  $\mathbf{\hat{w}}_{12}= \mathbf{\hat{u}}_1  \times \mathbf{\hat{u}}_2 /  \left | 
\mathbf{\hat{u}}_1 \times \mathbf{\hat{u}}_2 \right |$.
$H_{22}$ vanishes when the two helices are parallel or perpendicular to each other ($\beta_{12}$ is an integer multiple of $90^\circ$), and takes the largest absolute values for $\beta_{12}$ equal to odd multiples of $45^\circ$;
it is positive for $\cal R$- and and negative for $\cal L$-skewed configurations (see Fig. \ref{fig:descriptor}). 
After integration of $H_{22}$ over the whole excluded volume, the result is 
$h_{22}>0$ if $\cal R$ pair configurations have an excluded volume larger than $\cal L$ configurations;
$h_{22}<0$ in the converse case.

\section{Results}
\noindent We have performed calculations for model helices made of fused hard spheres.\cite{JCPI,JCPII} An example is shown in Fig. \ref{fig:helix}, where also the geometric parameters of a helix are shown: the radius $r$, the pitch $p$, the contour length $L$ and the Euclidean length $\Lambda$. All calculations were performed for right-handed helices, using the same numerical procedure  presented in ref. \cite{Tombolato05}: at a given density the chiral strength $k_2$ and the twist elastic constant $K_{22}$, and then the cholesteric pitch $\cal P$,  are calculated in terms of the orientational order parameters,  $\braket{P_{j}}$,  which are obtained by minimization of the free energy of the undeformed nematic phase.
Results will be reported in reduced units, with the diameter $D$ taken as the unit of length;
moreover, scaled values of the chiral strength, $k_2^\ast=k_2 /k_{B}T$, and of the twist elastic constant, $K_{22}^\ast=K_{22}/k_{B}T$, will be reported.

\begin{figure}[h]
\begin{center}
\includegraphics[scale=0.9]{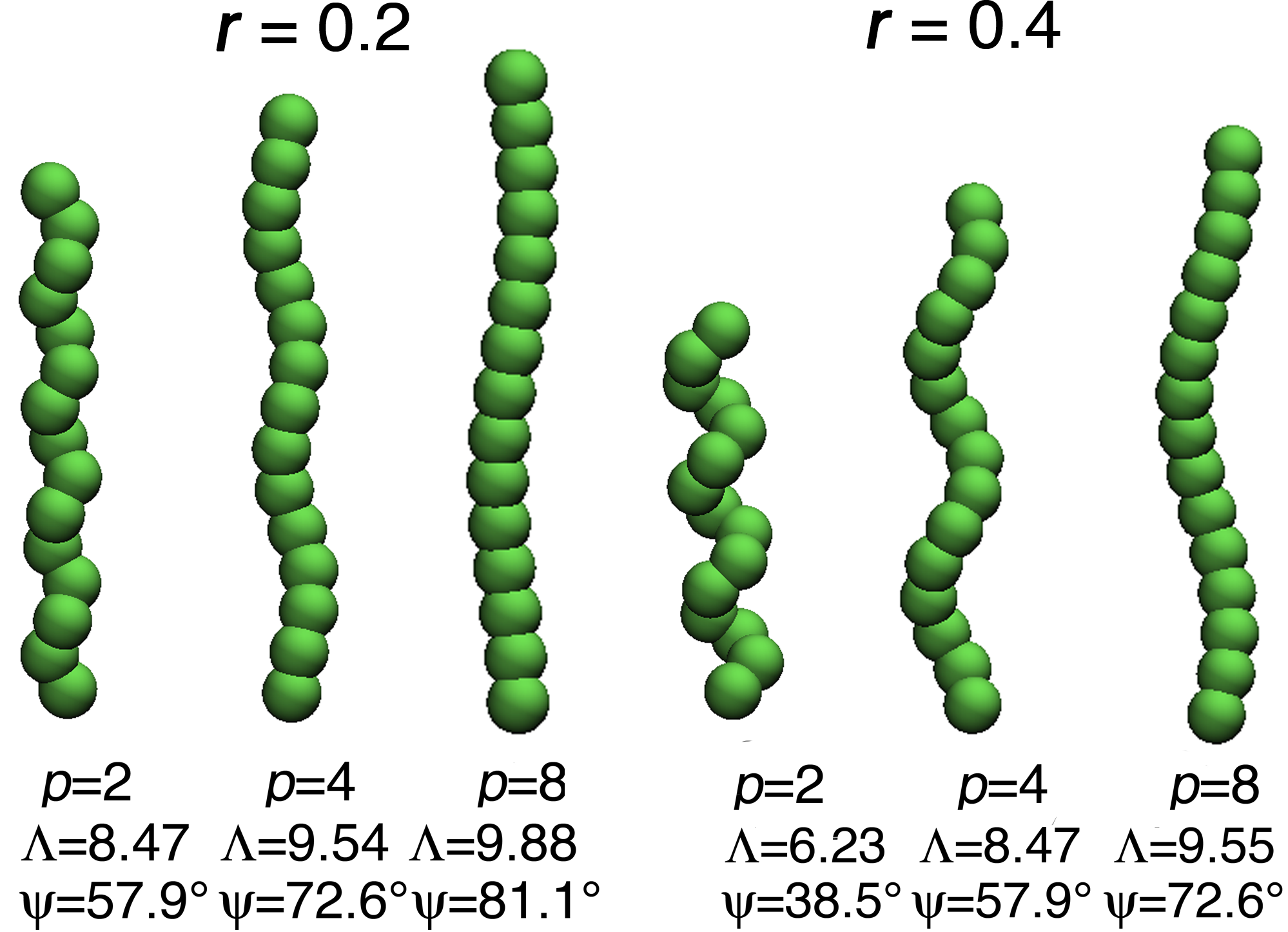}
\caption {Helices having the same contour length, $L=10$, and different values of the radius $r$ and pitch $p$. Under each helix the inclination angle $\psi$ and the Eucledian length $\Lambda$ are reported.}
\label{fig:hel_2}
\end{center}
\end{figure}

Firstly, we have performed calculations for the right-handed helices shown in Fig. \ref{fig:hel_2}, whose phase diagram was investigated in refs. \cite{JCPI,JCPII}: they have constant contour length $L=10$, and different values of the radius $r$ and pitch $p$.
Table \ref{tab:k2helices} reports the 
$h_{jj}$ pseudoscalars, defined by Eq. \eqref{eq:hii}, calculated for these systems.
For all helices positive $h_{22}$ values are obtained.
The higher rank pseudoscalars, $h_{44}$ and $h_{66}$, 
may be positive or negative, and the sign seems to depend on the shape details. 
The simultaneous presence of contributions of opposite sign for helices with a given handedness 
is not surprising, since analogous observations have already been reported,
\cite{Tombolato05,SpadaChemEurJ2006} and can be explained in the following way.
A variety of pair configurations of helices are possible and, of 
 the $\cal R$ and $\cal L$-skewed with the same twist angle, in some cases the former, in some cases the latter gives smaller excluded volume. For the majority of the helices reported in Table \ref{tab:k2helices}, the presence of an $h_{jj}$ term that is positive and much larger than the others indicates that, altogether,  
 $\cal R$-skewed pair configurations lead to a larger excluded volume than $\cal L$-skewed configurations.  We can observe that such cases correspond to inclination angles larger than $40^\circ$ (see Fig. \ref{fig:hel_2}) and intrinsic pitches $p$ sufficiently larger than the bead diameter $D$.

The wide range of $h_{22}$ values reported in Table  \ref{tab:k2helices} indicates that the difference in excluded volume strongly depends on the helix morphology.
As a general rule, higher  differences are obtained for helices with larger $p$ and $r$ values. 
This is well illustrated by the cases $r=0.2$, $p=2$, and $r=0.4$, $p=4$, which are characterized by the same inclination angle: $h_{22}$ increases by an order of magnitude on going from the former to the latter.
This behavior can be easily understood considering that larger pitch and radius imply deeper grooves and  more marked deviations from a cylindrical shape. 

\begin{table}[h]
\small
\caption{Pseudoscalars of different ranks, $h_{jj}$, 
calculated for helices of the same contour length $L=10$, and 
different pitch $p$ and radius $r$ (see Fig. \ref{fig:hel_2}). Lengths are scaled with the bead diameter $D$.}
\begin{tabular*}{0.5\textwidth}{@{\extracolsep{\fill}}llllll}
\hline
& & $h_{22}$ &  $h_{44}$ & $h_{66}$ \\ 
\hline
$r=0.2$ & $p=2$ &   15.2  &  -9.0    &  -1.4    \\ 
           & $p=4$ &       78.5  &  -0.7    &   -5.6  \\
           & $p=8$ &       66.1  &  12.6  &  2.7      \\
$r=0.4$ & $p=2$ &      2.4   & -8.4   &  0.8       \\
            & $p=4$ &  147.4   & -29.0  &  -9.6       \\
            &  $p=8$ & 212.3   & 23.4  &  -1.7        \\
\hline
 \end{tabular*}
\label{tab:k2helices}
\end{table}

Table \ref{tab:k2K22tot} shows the 
scaled chiral strength $k^\ast_2$ and twist elastic constant $K^\ast_{22}$, along with the $\braket{P_2}$ order parameter and the cholesteric pitch $\cal P$ calculated for the helices in Fig. \ref{fig:hel_2}.  
The data refer to the density of the cholesteric-to-isotropic (N$^\ast$-I) phase transition. 
We can see a clear correlation between chiral strength and $h_{22}$ values: positive $k^\ast_2$ is predicted for all helices but the one with $p=2$,  $r=0.4$, for which a small negative  $k^\ast_2$ is obtained.  
Also the chiral strengths cover a wide range and, for a given intrinsic pitch, comparatively 
higher values are predicted for the helices with $r=0.4$, which are more curled than those with a smaller radius.
This does not occur for $p=2$, because in this case strong interlocking of pair of helices is not possible, due to the small size of the helix pitch compared to the diameter of beads.

The data reported in Table \ref{tab:k2K22tot} show that the twist elastic constants are much less sensitive than chiral strengths to changes in the helix structure: their variation along the series of helices under examination does not exceed 20$\%$.  Like the $\braket{P_2}$ order parameters (see ref. \cite{JCPI}  for the discussion of the transition properties of hard helices), they are mainly determined by the aspect ratio of particles. 
 
Considering now the cholesteric pitch $\cal P$,  we can see that all helices, 
with the exception of the case $r=0.4$, $p=2$, are found to form $\cal L$ cholesteric phases. 
As a consequence of the relatively small differences in $K_{22}$ values, the magnitude of the cholesteric pitch reflects the trend of $k_2$, with the tighter $\cal P$ values for the curlier helices.  
It is worth pointing out that the cholesteric pitch predicted for helices with $r=0.4$ and $p > 2$, which is less than 100 times the molecular size, is remarkably small. Typical experimental values for lyotropic liquid crystals are at least one order of magnitude higher.\cite{RobinsonTetra61,PBLG,LivolantDNA,FradenPRL03}

Interestingly,  we have found that in the case $r = 0.2$, $p = 2$ the
sign of the calculated cholesteric pitch changes with density:
at the N$^\ast$-I transition $\cal P$ is large and negative ($\cal L$ cholesteric),
and then increases in magnitude with increasing density until,
at a certain point, it reverses. This behaviour reflects a change
in the relative contribution of the terms of various ranks in
Eq. \ref{eq:k2}, due to a different density dependence of the $\braket{P_j}$ order parameters.
From the physical point of view this corresponds to
a system where, as density increases, the cholesteric helix first
unwinds, and then rewinds in the opposite sense, passing
through an untwisted nematic phase. In the literature, there
are both experimental and theoretical examples of cholesteric
inversion as a function of temperature or density (a review can be
found in ref. \cite{KatsonisJMaterChem}), which were ascribed to the competition of
factors promoting twists in opposite senses: examples of such
factors are inter-particle interactions of different nature or
different molecular conformations. More subtle effects have been
related to a different temperature dependence of the orientational
order parameters for different molecular axes, in the case
of biaxial particles.\cite{FMN96,Osipov2000}  This has an analogy with the behavior just
discussed for the hard rigid helices with  $r = 0.2$, $p = 2$.

\begin{table}[h]
\small
\caption{Scaled chiral strength $k^{\ast}_2$ and twist elastic constant  $K^\ast_{22}$, $\braket{P_2}$  order parameter and  cholesteric pitch $\cal P$, for helices of length $L=10$  and different values of the pitch $p$ and radius $r$ (see Fig. \ref{fig:hel_2}).  
Data are for state points in the N$^\ast$ phase at the the N$^\ast$-I phase transition.  Lengths are scaled with the sphere diameter $D$.}
  \begin{tabular*}{0.5\textwidth}{@{\extracolsep{\fill}}lllllll}
\hline
& & $k^{\ast}_2 \cdot 10^{4}$ &  $K^{\ast}_{22} $ & $\braket{P_2}$ & $\cal P$ \\
\hline
$r=0.2$ & $p=2$ &  4.83   &   0.154 & 0.64 &  -2008   \\
           & $p=4$ &  41.36  &    0.177  & 0.66 &  -268   \\
           & $p=8$ &   29.03 &    0.184    & 0.68 &  -399 \\
$r=0.4$ & $p=2$ &    -3.83 &  0.153 & 0.60 &  2509 \\
            & $p=4$ &   98.35 &   0.152  & 0.61 &   -97  \\
            &  $p=8$ &  110.13  & 0.159   & 0.61   &   -90 \\
\hline
 \end{tabular*}
\label{tab:k2K22tot}
\end{table}

\begin{figure}
   \centering
   \includegraphics[scale=0.8]{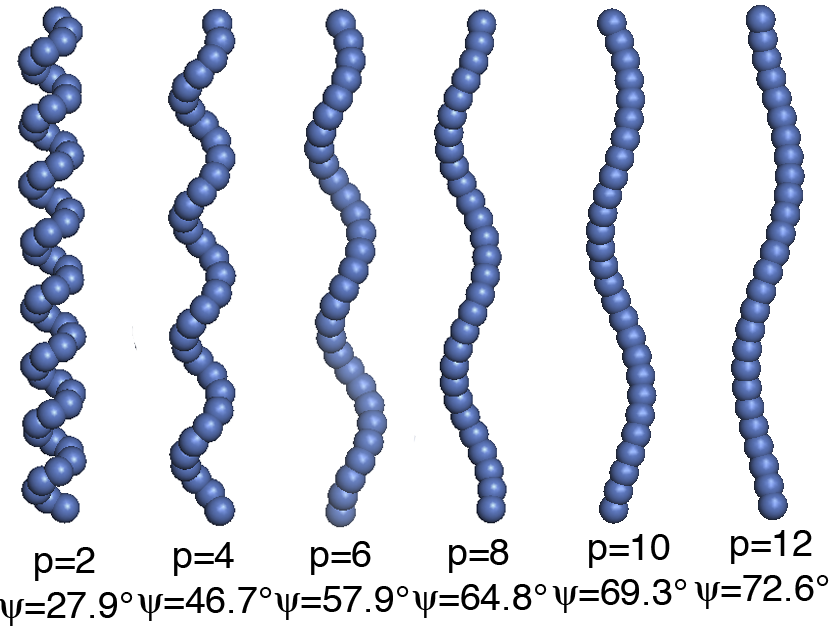} 
   \caption {Helices having the same Euclidean length, $\Lambda=16$, and radius, $r=0.6$, but different values of the pitch $p$. Under each helix the inclination angle $\psi$ is reported.}
   \label{fig:helicesr06t}
\end{figure}

To analyse more systematically  the relationship between inclination angle and cholesteric handedness we have considered  a set of helices having the same  Euclidean length ($\Lambda=16$) and radius ($r=0.6$), with pitch $p$ equal to 2, 4, 6, 8, 10 and 12.  
Along this series the inclination angle is only a function of the intrinsic pitch and gradually changes from $\sim 28^\circ$ to $\sim 72.5^\circ$
(see Fig. \ref{fig:helicesr06t}).

\begin{figure}
 \centering
\includegraphics[scale=0.7]{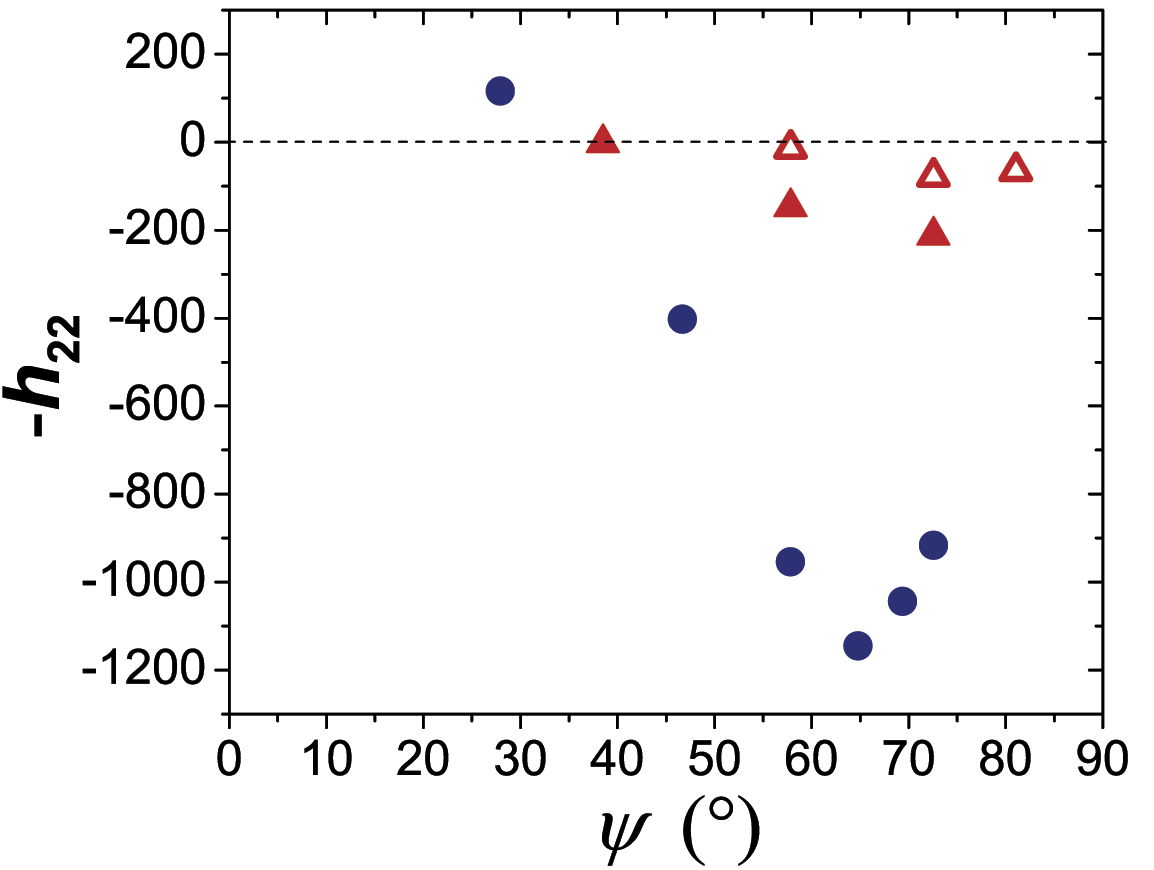}
\caption{Opposite of the second rank pseudoscalar $h_{22}$ as a function of the inclination angle $\psi$, for different helices (all right-handed). Blue circles: helices with $\Lambda=16 D$, $r=0.6$ and $p=2,4,6,8,10,12$. Closed triangles: helices with $L=10$, $r=0.4$, and $p=2, 4, 8$. Empty triangles: helices with $L=10$, $r=0.2$, and $p=2, 4, 8$. Lengths are scaled with the sphere diameter $D$.}
\label{fig:CPr06}
\end{figure}

Fig. \ref{fig:CPr06} shows the opposite of $h_{22}$ calculated for these helices as a function of the inclination angle. 
We can see  a  non-monotonic dependence upon the inclination angle (i.e. the intrinsic pitch of helices, here): 
for  $\psi \gtrsim 40^\circ$, $h_{22}$ is positive  and takes the highest values for $\psi$ around $65^\circ$. 
For  $\psi \lesssim 40^\circ$, $h_{22}$ becomes negative, but is small, which can be ascribed to the limited depth of the grooves 
(see Fig. \ref{fig:helicesr06t}), which 
prevents strong interpenetration of nearby helices and then, large differences in excluded volume between $\cal R$- and $\cal L$-skewed pair configurations.
For the sake of comparison we have reported in Fig. \ref{fig:CPr06}  
also the $h_{22}$  values  obtained for  the helices with constant contour length  shown in Fig. \ref{fig:hel_2}.
We can see that, irrespective of specific details, the dependence on the inclination angle exhibits a general trend.
The $h_{22}$  values obtained for the shorter helices are smaller, by a factor that simply depends on their size.

The helices in Fig. \ref{fig:helicesr06t}, having practically the same aspect ratio, are expected to have similar twist elastic constants; therefore the changes in $h_{22}$ can be related to those in cholesteric properties.
From the results reported in Fig. \ref{fig:CPr06} we can infer that those with $p>4$ will form $\cal L$ cholesteric phases, and the tightest pitch will be found for $p=8$. 
Only for the helix with $p=2$ we expect a long-pitch, $\cal L$-handed cholesteric phase.   

A trend analogous to that reported in Fig. \ref{fig:CPr06} was obtained also 
for hard-core interactions between DNA duplexes, in which case a coarse-grained representation of DNA was used and different values of $\psi$ were obtained by changing the twist angle between adjacent base pairs.\cite{RossiECLC2013} 
The canonical form of B-DNA has a right-handed helix, with inclination angle smaller than 30$^\circ$; in line with the present results, hard core interactions were predicted to lead to a right-handed cholesteric phase.\cite{Tombolato05,Frezza:SoftMatter}
Experimentally,  relatively long B-DNA ($\gtrsim 100$ base pairs) has been found to form a left-handed N$^\ast$ phase.\cite{LivolantDNA} However, both right- and left-handed organizations, depending on sequence, were found for DNA oligomers.\cite{ZanchettaPNAS} As an explanation of this behavior it was proposed that electrostatic interactions, superimposed to the hard-core ones,  would lead to inversion of the cholesteric twist for long B-DNA and some of the oligomeric systems.\cite{Tombolato05,Frezza:SoftMatter}

\subsection{Monte Carlo simulations of the isotropic phase}
\begin{figure}
 \centering
\includegraphics[scale=0.35]{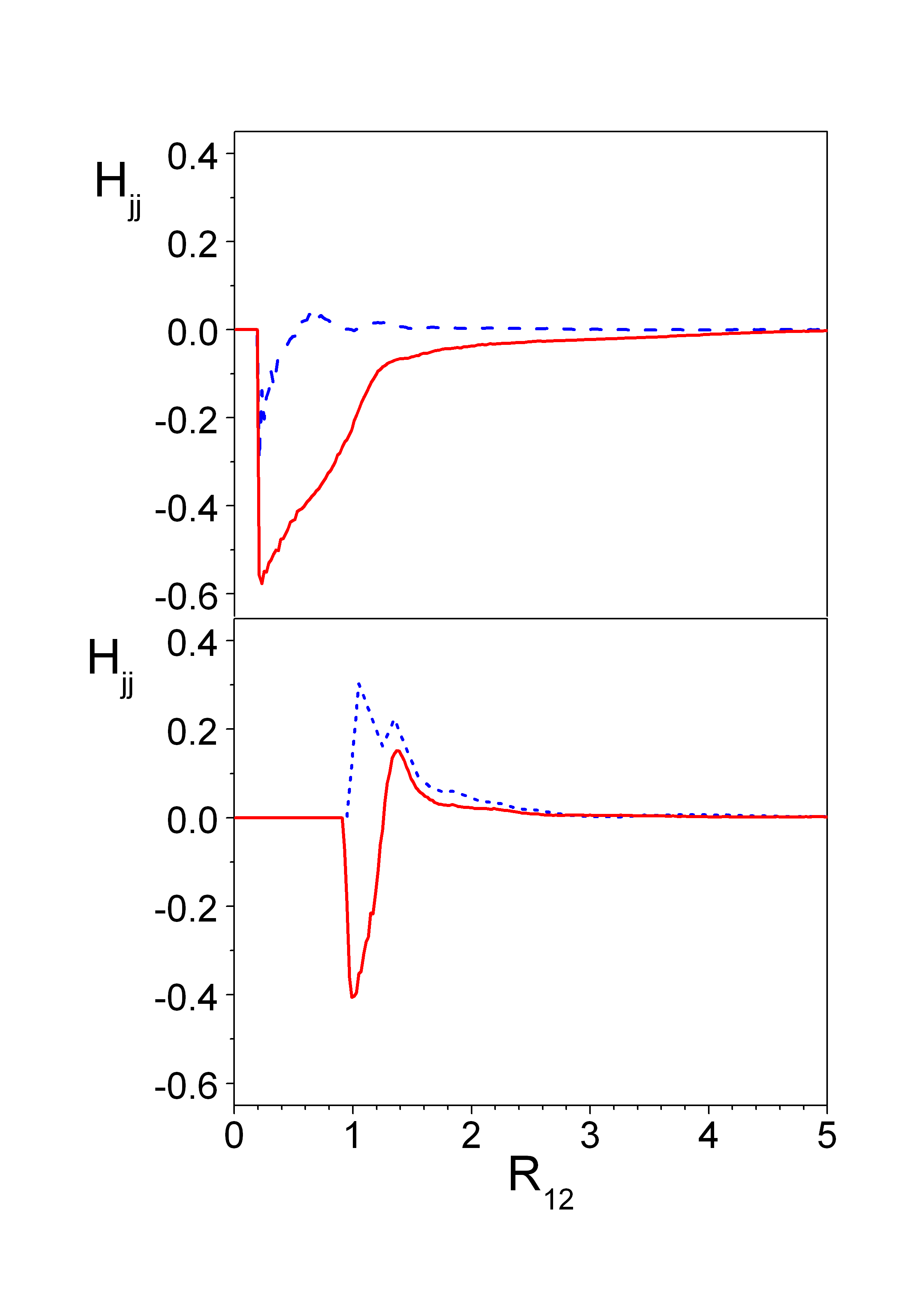}
\caption{Radial orientational correlations functions $H_{22}(R_{12})$ (solid) and  $H_{44}(R_{12})$ (dashed), as a function of the helix-helix distance $R_{12}$. Results from Monte Carlo simulations of helices with $L=10$, and $r=0.4$, $p=4$ (top) or $r=0.4$, $p=2$ (bottom). State points in the isotropic phase were taken, at the scaled pressure $P^\ast=0.6$ in the former, and $P^\ast=1.0$ in the latter case. Lengths are scaled with the bead diameter $D$.}
\label{fig:MC}
\end{figure}

\noindent The predictions of Onsager-like theory can be substantiated by the analysis of Monte Carlo trajectories.
Fig. \ref{fig:MC} shows results obtained from simulations for  helices of contour length $L=10$, with $r=0.4$, $p=2$ and $r=0.4$, $p=4$.
The quantities shown in the plots the are radial orientational correlation functions $H_{22}(R_{12})$ and  $H_{44}(R_{12})$, with $H_{jj}$ defined as in Eqs. \ref{eq:subeqns}, as a function of the inter-particle distance $R_{12}$. 
These correlation functions correspond, apart from a multiplicative factor, to those denoted as  $S_{jj1}$ in ref. \cite{Memmer2000}   
Simulations in the $NPT$ ensemble were carried out, using the same procedure outlined in refs. \cite{JCPI,JCPII}. 
State points in the isotropic phase were taken, at  the scaled pressure $P^\ast=PD^3/k_BT=1.0$ ($\rho=0.048105$) in the case $r=0.4$, $p=2$, and $P^\ast=0.6$ ($\rho=0.041066$) for $r=0.4$, $p=4$.

The functions $H_{22}(R_{12})$ and $H_{44}(R_{12})$ account for chiral correlations between the $\mathbf{\hat{u}}$ axes of helices. We can see in Fig. \ref{fig:MC} that  they are equal to zero almost everywhere, but  at very short distances, close to the diameter of a bead, they take non-vanishing values. This indicates the absence of long range chiral correlations, which is expected in the isotropic fluid phase. However the peaks that appear at very close distances demonstrate the preference for a given skewness of the long axes of helices. 
Such peaks would be absent in the case of achiral particles.  

Going more into details, we can notice some differences between the results obtained for the two kinds of helices. For the case $r=0.4$, $p=4$, non-vanishing values of $H_{jj}$  fall in a range of distances shorter than the bead radius. This can be understood considering that helices are non-convex bodies and their center, used to define inter-particle distances, was taken in the middle point of their axis. Therefore, values  $R_{12}<D$ correspond to interlocked helices. 
$H_{22}$ is throughout negative and much larger than $H_{44}$, which can take both positive and negative values.
So, we can infer that the long axes of pairs of helices at the contact distances are preferentially found in $\cal L$-skewed configurations.
This is in line with the results reported in table \ref{tab:k2helices}  for the helices with $r=0.4$, $p=4$ (it may be worth recalling that positive/negative $h_{jj}$ values indicate that $\cal L$/$\cal R$-skewed configurations lead to smaller excluded volume).

For helices with $r=0.4$, $p=2$, the non-vanishing values of both $H_{jj}$ functions are shifted towards slightly longer distances ( $R_{12} \sim D$), because the particle morphology prevents deep interlocking. Unlike the previous case, here we cannot easily identify the preference for a given skewness.
$H_{22}$ features both $\cal L$ and  $\cal R$ contributions, comparable in magnitude, whereas $H_{44}$ is positive, indicating a preference for $\cal R$-skewed configurations.
Again, we can recognize a correspondence with the $h_{jj}$ values reported
in Table \ref{tab:k2helices} for the helix $r=0.4$, $p=2$. 

\section{Conclusions}
\noindent Using an Onsager-like theory we have investigated the relationship between the handedness and pitch of hard helices and those of the cholesteric phase that they form.   
For the N$^\ast$ handedness, our numerical results are in broad agreement with Straley arguments for threaded rods: \cite{Straley}  taking  the inclination angle of the helical motif as a useful descriptor,
we have found that right-handed helices  form $\cal L$ cholesterics if  
$\psi \gtrsim 40^\circ$, and $\cal R$ cholesterics if $\psi \lesssim 40^\circ$. 
For left-handed helices the opposite would occur.  
This can be seen as  a general behavior for helices with hard-core interactions, provided that their  
pitch $p$ is larger than the  thickness of the helical protrusions  (in our case, for $p > D$). 
According to our calculations, helices with $\psi$ around $40^\circ$ belong to a gray zone, where the cholesteric handedness may depend on details of the morphology. Anyway, they are the less effective in twisting the director and are predicted to lead to cholesteric phases with a long pitch, which may even  change its handedness as a function of density. 
As for the magnitude of the cholesteric pitch, we have found that it is  a non-monotonic function of the inclinations angle  $\psi$ and, comparing helices with the same aspect ratio, which then have similar twist elastic constant, the tightest pitches are predicted for $\psi \sim 65^\circ$.

The driving force behind the formation of the cholesteric phase by hard particles is the entropy gain deriving from the director twist. Within the Onsager-like theory, this is traced back to the difference in 
excluded volume between configurations of pairs of helices having $\cal R$ and $\cal L$-skewed axes.  
Our Onsager-like approach leads to the definition of a hierarchy of pseudoscalars ($h_{jj}$),
related to rotational invariant functions,\cite{Stone78}
which quantify these differences.
For helices with $\psi$ either sufficiently larger or smaller than $40^\circ$, the single lowest rank term, $h_{22}$, prevails over the others,
and can be roughly related to the inverse cholesteric pitch. On the other hand, when $\psi \sim 40^\circ$ or $p \sim D$  
the differences in excluded volume between $\cal R$ and $\cal L$ pair configurations  are relatively small, and their 
net balance may be  a subtle function of density.
This is reflected by $h_{jj}$ contributions of different rank similar in magnitude and different in sign.  

We have found that the results of the Onsager-like theory compare very well with those extracted from Monte Carlo simulations of hard helices in the isotropic phase.  In principle, the cholesteric pitch could be obtained from numerical simulations of the N$^\ast$ phase. 
However this remains a challenging task, which poses a difficulty in selecting and then handling suitable periodic boundary conditions,\cite{Allenmasters,Memmer2000, Schoen2013} especially for systems of freely translating and rotating particles interacting with sole hard-body interactions.
Simulations of the isotropic phase do not give evidence of long-range  chiral correlations, but  we have shown that  they  contain signatures of short-range chiral correlations between the axes of helices. In the isotropic phase such correlations are lost within a few molecular lengths, but liquid crystal ordering allows them to propagate to a much longer scale. 

Interestingly, for strongly curled hard helices we predict very tight cholesteric pitches, of the order of one hundred of times the characteristic size of particles. There are no evidences of such strong distortions in real systems: typical values of the cholesteric pitch for lyotropic liquid crystals are at least one order of magnitude higher.\cite{KatsonisJMaterChem}  
One reason is that in real systems there are also other interactions, which can compete with steric repulsions.\cite{FradenPRL03,DogicSoftMatter,ProniChemEurJ} Moreover helical macromolecules are generally endowed by some flexibility that, besides affecting the phase diagram,\cite{DeGaetani08} has the effect of reducing the net chirality.\cite{KatsonisJMaterChem}  
The ideal systems for direct comparison with our prediction are cholesteric suspensions of rigid colloidal particles, as could be obtained by the synthesis techniques nowadays available.\cite{FischerNM}
One last point has to be remarked: the results reported here were obtained  for the hypotetical N$^\ast$ phase that beyond a certain packing fraction  becomes more stable than the isotropic phase. However,  under the same conditions there might be other competing phases, which have not been considered here.
Indeed, we have found that the width of the cholesteric range decreases with the curliness of helices, and strongly curled helices undergo a direct transition from the isotropic to the screw-like nematic phase.\cite{JCPII} The relationship between cholesteric and screw-like order is an open very interesting issue, which we intend to address in future studies.

\section{Note}
\noindent We recently became aware of another study of the cholesteric
phase formed by hard helices.\cite{Belli2014} A different method was used to
calculate the cholesteric pitch, but the results are consistent
with those reported here.

\begin{acknowledgments}
\noindent H.B.K., E.F., A.F. and A.G. gratefully acknowledge support from PRIN-MIUR 2010-2011 project (contract 2010LKE4CC). 
G.C. is grateful to the Government of Spain for the award of a Ram\'{o}n y Cajal research fellowship.
\end{acknowledgments}


\begin{thebibliography}{99}

\bibitem{JCPI}
E. Frezza, A. Ferrarini, H.B. Kolli, A. Giacometti, G. Cinacchi,
\textit{J. Chem. Phys.} \textbf{138}, 164906 (2013).

\bibitem{JCPII}
H.B. Kolli, E. Frezza, G. Cinacchi, A. Ferrarini, A. Giacometti, T.S. Hudson,
\textit{J. Chem. Phys.}  \textbf{140}, 081101 (2014).

\bibitem{Barry06}
E. Barry, Z. Hensel, Z. Dogic, M. Shribak and R. Oldenbourg, \textit{Phys. Rev. Lett.} \textbf{96}, 018305 (2006).

\bibitem{Manna07}
F. Manna, V. Lorman, R. Podgornik, and B. Zeks, Phys. Rev. E \textbf{75}, 030901(R) (2007).

\bibitem{note}
Unlike most literature, here `chiral nematic' will not be used as a synonym of cholesteric, since this could be ambiguous after the identification of other chiral nematic phases such as, in the case of helical particles, the screw-like nematic (N$^\ast_s$).\cite{Barry06,Manna07,JCPII}

\bibitem{Straley}
J.P. Straley, \emph{Phys. Rev. A} \textbf{14}, 1835 (1976).

\bibitem{Harris1999}
A. B. Harris, R.D. Kamien, and T. C. Lubensky, \emph{Rev. Mod. Phys.} \textbf{71}, 1745 (1999).

\bibitem{Cherstvy2008}
A. Cherstvy, \emph{J. Phys. Chem. B} \textbf{112}, 12585 (2008).

\bibitem{Pelcovits1996}
R. A. Pelcovits, \emph{Liq. Cryst.} \textbf{21}, 361 (1996).


\bibitem{Evans1992}n
G. T. Evans,  \emph{Mol. Phys.} \textbf{77}, 969(1992).

\bibitem{Harris1997}
A. B. Harris, R.D. Kamien, and T. C. Lubensky, \emph{Phys. Rev. Lett.}  \textbf{78}, 1476 (1997).

\bibitem{JacksonJMolPhys2011}
S. Varga and G. Jackson, \emph{Mol. Phys.} \textbf{109}, 1313 (2011).

\bibitem{KatsonisJMaterChem}
N. Katsonis, E. Lacaze, and A. Ferrarini, \emph{J. Mater. Chem.} \textbf{22}, 7088 (2012).

\bibitem{Pieraccini11}
S. Pieraccini, S. Masiero, A. Ferrarini and G. P. Spada, \emph{Chem. Soc. Rev.}  \textbf{40}, 258 (2011).

\bibitem{Sato1998}
T. Sato, J. Nakamura, A. Teramoto and M. M. Green, \emph{Macromolecules} \textbf{104}, 6755 (1996).

\bibitem{LivolantDNA}
F. Livolant and A. Leforestier, \emph{Prog. Polym. Sci.} \textbf{21}, 1115 (1996).

\bibitem{Zugenmaier}
P. Zugenmaier, in \emph{Handbook of Liquid Crystals}, ed.
D. Demus, J. Goodby, G. W. Gray, H.-W. Spiess and V. Vill,
Whiley-WCH, 1998, vol. 4.

\bibitem{wensink_JPhysCondensMatter}
H. H. Wensink and G. Jackson, \emph{J. Phys.: Condens. Matter}  \textbf{23}, 194107 (2011).


\bibitem{wensink:234911}
H. H. Wensink and G. Jackson, \emph{J. Chem. Phys.}  \textbf{130}, 234911 (2009).

\bibitem{Osipov85}
M. A. Osipov, \emph{Chem. Phys.} \textbf{96}, 259 (1985).


\bibitem{Emelyanenko2003}
A. V. Emelyanenko, \emph{Phys. Rev. E} \textbf{67}, 031704 (2003).



\bibitem{Tombolato05}
F. Tombolato, and A. Ferrarini, \emph{J. Chem. Phys.} \textbf{122}, 054908 (2005).

\bibitem{Frezza:SoftMatter}
E. Frezza, F. Tombolato, and A. Ferrarini, \emph{Soft Matter} \textbf{7}, 9291 (2011).


\bibitem{McQuarrie}
D. A. McQuarrie, \textit{Statistical Mechanics} (University Science Books, Sausalito, CA, 2000).


\bibitem{Varga2000}
S. Varga and I. Szalai, \emph{Mol. Phys.} \textbf{98}, 693 (2000).

\bibitem{Vertogen}
G. W. Vertogen and W. de Jeu,  \textit{Thermotropic Liquid Crystals. Fundamentals} (Springer, Berlin, 1998).

\bibitem{Varshalovich}
A. D. Varshalovich, N. A. Moskalev, and V. Kersonskii,  \textit{Quantum Theory of Angular Momentum} (World Scientic, NewYork, 1995).

\bibitem{Stone78}
A. J. Stone, \emph{Mol. Phys.} \textbf{36}, 241 (1978).

\bibitem{RossiECLC2013}
M. Rossi, E. Frezza, G. Zanchetta, A. Ferrarini, and T. Bellini (Presented at the 12th European Conference on Liquid Crystals, Rhodes, Greece, 2013).


\bibitem{FrezzaPhD}
E. Frezza, Ph.D. thesis, Universit\`{a} di Padova (2014).

\bibitem{SpadaChemEurJ2006}
S. Pieraccini, A. Ferrarini, K. Fuji, G. Gottarelli, S. Lena, K. Tsubaki and G. P. Spada, \emph{Chem. - Eur. J.} \textbf{12}, 1121 (2006).

\bibitem{RobinsonTetra61}
C. Robinson, \emph{Tetraheron} \textbf{13}, 219 (1961).

\bibitem{PBLG}
C. Robinson, \emph{Trans. Faraday Soc.} \textbf{52}, 571 (1956).


\bibitem{FradenPRL03}
E. Grelet and S. Fraden, \emph{Phys. Rev. Lett.} \textbf{90}, 198302 (2003).

\bibitem{FMN96}
A. Ferrarini, G. J. Moro and P. L. Nordio, \emph{Mol. Phys.} \textbf{87}, 485 (1996).

\bibitem{Osipov2000}
A. V. Emelyanenko, M. A. Osipov and D. A. Dunmur, \emph{Phys. Rev. E: Stat. Phys., Plasmas, Fluids, Relat. Interdiscip. Top.} \textbf{62}, 2340 (2000).


\bibitem{ZanchettaPNAS}
G. Zanchetta, F. Giavazzi, M. Nakata, M. Buscaglia, R. Cerbino, N.A. Clark, and T. Bellini,  \emph{Proc. Natl. Acad. Sci } \textbf{107}, 17497 (2010).

\bibitem{Memmer2000}
R. Memmer, \emph{Liq. Cryst.} \textbf{27}, 533 (2000).

\bibitem{Allenmasters}
M. P. Allen and A. J. Masters, \emph{Mol. Phys.} \textbf{79}, 277 (1993).

\bibitem{Schoen2013}
M. Melle, M. Theile, C. H. Hall and M. Schoen, \emph{Int. J. Mol. Sci.} \textbf{14}, 17584 (2013).

\bibitem{DogicSoftMatter}
 E. Barry, D. Beller, and Z. Dogic, \emph{Soft Matter} \textbf{5}, 2563 (2009).
 
\bibitem{ProniChemEurJ}
G. Proni, G. Gottarelli, P. Mariani, and G. Spada, \emph{Chem.-Eur. J.} \textbf{6}, 3249 (2000).


\bibitem{DeGaetani08}
G. Cinacchi and L. De Gaetani, \emph{Phys. Rev. E: Stat. Phys., Plasmas, Fluids, Relat. Interdiscip. Top.} \textbf{77}, 051705 (2008).



\bibitem{FischerNM}
A. G. Mark, J. G. Gibbs, T.-C. Lee and P. Fischer, \emph{Nat. Mater.} \textbf{12}, 802 (2013).


\bibitem{Belli2014}
S. Belli et al., arXiv:1404.2113.






\end{thebibliography}
\end{document}